\documentclass[
 aip,
 amsmath,amssymb,
 reprint,%
]{revtex4-1}

\usepackage{graphicx}% Include figure files
\usepackage{dcolumn}% Align table columns on decimal point
\usepackage{bm}% bold math

\usepackage[utf8]{inputenc}
\usepackage[T1]{fontenc}
\usepackage{mathptmx}

\usepackage[english]{babel}

\begin{document}

\preprint{AIP/123-QED}

\title[]{The effect of rigid electron rotation on the Grad-Shafranov equilibria of a class of FRC devices}

\author{C.P.S. Swanson}
%\author{S.J. Thomas}
\affiliation{Princeton Fusion Systems, Plainsboro, NJ, USA}
\author{S.A. Cohen}
\affiliation{Princeton Plasma Physics Laboratory, Princeton, NJ, USA}

\date{\today}% It is always \today, today,

\begin{abstract}
Rigid electron rotation of a fully penetrated Rotamak-FRC produces a pressure flux function that is more peaked than the Solov'ev flux function. This paper explores the implications of this peaked pressure flux function, including the isothermal case, which appear when the temperature profile is broader than the density profile, creating both benefits and challenges to a Rotamak-FRC based fusion reactor. In this regime, the density distribution becomes very peaked, enhancing the fusion power. The separatrix has a tendency to become oblate, which can be mitigated by flux conserving current loops. Plasma extends outside the separatrix, notably in the open field line region. This model does not apply to very kinetic FRCs or FRCs in which there are significant ion flows, but it may have some applicability to their outer layers.
\end{abstract}

\maketitle

\section{Introduction}

Rotamak-FRCs are plasma physics experiments in which electron current is driven by an externally-imposed rotating magnetic field (RMF).\cite{hoffman_principal_2006} Interest in this configuration of plasma arises from its favorable properties for scaling into a nuclear fusion reactor, particularly a compact one.\cite{cohen_direct_2019,cohen_reducing_2015,slough_FRC_1999,miller_overview_1998} The favorable properties include: High plasma $\beta$, maximizing the plasma pressure for a given magnetic field; low internal field, allowing high-temperature and advanced fuels; and a simple, compact, and efficient method of heating and current drive in the form of the RMF system. RMF current drive dates to the 1960s.\cite{blevin_plasma_1962} There were several Rotamak-FRC experiments operating in the 1990s and 2000s.\cite{miller_overview_1998,jones_experimental_1987,hoffman_TCS_2002,jones_review_1999,cohen_formation_2007} An existing example is the Princeton Field-Reversed Configuration 2 (PFRC-2) experiment at the Princeton Plasma Physics Laboratory (PPPL).\cite{cohen_RF_2011}

In fully penetrated Rotamak-FRCs, the current drive is assumed to be due to electrons rotating in synchrony with the applied external RMF while the ions are stationary.\cite{hugrass_production_1979,hugrass_numerical_1981,jones_steady-state_1981,jones_review_1999,milroy_numerical_1999,hoffman_principal_2006} There is also speculation that FRCs that are not driven by RMF will also rotate synchronously due to collisional effects.\cite{rostoker_equilibrium_2002} RMF-synchronous electron rotation has been observed in experiments\cite{guo_rotating_2007} and PIC (kinetic) simulation.\cite{welch_formation_2010}

A simplified (no $B_t$) Grad-Shafranov model is often used to predict and reconstruct the MHD equilibria of these plasma configurations.\cite{grad_hydromagnetic_1958,solovev_plasma_1970} While many modern analyses assume a Solov'ev\cite{solovev_plasma_1970} pressure flux function, assuming $P\propto \Psi$ (pressure linear in flux),\cite{petrov_experiments_2010,euripides_rotamak_1997,bellan_particle_1989} it has been known since 1982 that the Solov'ev linear pressure flux function is the \textit{least} steep pressure flux function consistent with rigid rotation, and that more realistic flux functions have a higher power law, $P\propto \Psi^N$, $N\geq1$\cite{storer_pressure_1982,storer_compact_1983,donnelly_magnetohydrodynamic_1987} or even an exponential relationship, $P\propto e^{\Psi/\Psi_0}$.\cite{christofilos_astron_1958,marder_bifurcation_1970,armstrong_fieldreversed_1981,belova_numerical_2000,rostoker_equilibrium_2002,gota_separatrix_2003} The well-known Rigid Rotor 1-D radial pressure profile $P(r),B(r)$ implicitly assumes this pressure flux function. 

Where these steeper pressure flux functions have been used to numerically generate equilibria, it has predominantly been for the purpose of fitting to experimental measurements. In this context, the implications of this steeper pressure flux function on fusion reactor design have not been explored in detail. 

The Grad-Shafranov equation has also been used to model FRC equilibria with pressure flux functions that do not include rigid rotor effects.\cite{steinhauer_two-dimensional_2014,hewett_twodimensional_1983}

In Section \ref{sec:theory}, we will derive the flux functions to be inserted into the Grad-Shafranov solver that are required by the condition of rigid electron rotation. In Section \ref{sec:likelymu}, we will discuss the likely values of the free parameters that are defined in Section \ref{sec:theory}, the likely relative peakedness of the density and temperature profiles. In Section \ref{sec:applicability}, we will discuss the applicability of this model to experiments and reactors, and sketch alterations that may be required. In Section \ref{sec:solver}, we discuss the solver which produces self-consistent MHD equilibria from the equations in Section \ref{sec:theory}. In Section \ref{sec:results}, we will discuss the results of these MHD equilibria. In Section \ref{sec:conclusion}, we will conclude with a discussion of these results and their effect on the future design of Rotamak-FRC based fusion reactors.

\section{The pressure flux functions produced by rigid rotation}
\label{sec:theory}

In this section we will derive the pressure flux function to substitute into the Grad-Shafranov equation.\cite{grad_hydromagnetic_1958,solovev_plasma_1970} The Grad-Shafranov model has the pressure flux function as a free parameter, but as we will see here, the condition of rigid electron rotation implies a functional form.

The central assumption of this model is that RMF-driven current (Equation \ref{eq:jfromn}) is the diamagnetic current (Equation \ref{eq:mhd}), that is that $\vec{j}\times\vec{B}$ balances $\vec{\nabla}P$. The net effect of RMF is that it forces density to migrate across field lines until the pressure profile is such that the diamagnetic electron velocity rotates synchronously. This model also includes an isotropic pressure and no ion flows. This model is of limited applicability to reactor concepts in which kinetic effects dominate and ion flows are strong.

\subsection{Diamagnetic current condition, MHD equilibrium}

We assume an ideal axisymmetric MHD equilibrium, and that the plasma pressure at any given point is a function only of the enclosed flux, $P(\Psi(r,z))$. The equilibrium equation is simplified when it is assumed that there is no toroidal magnetic field.

The equilibrium condition is
\begin{equation}
\vec{j}\times\vec{B} = \vec{\nabla}P
\label{eq:mhd}
\end{equation}

By assuming axisymmetry, no toroidal field, and an isotropic pressure flux function $P(\Psi)$, this equation becomes:
\begin{equation}
j_\phi =2\pi r \partial_\Psi P
\label{eq:jfromp}
\end{equation}
where $\phi$ is the toroidal or azimuthal direction, $P=\Sigma nT$ is the plasma pressure, $r$ is the radial coordinate, $\Psi(r,z) = \int_0^r dr^\prime 2\pi r B_z(r^\prime,z)$ is the enclosed flux at point $(r,z)$, $\partial_\Psi P$ is the derivative of the plasma pressure $P$ with respect to the magnetic flux $\Psi$. $\Psi$ has not been normalized by $2\pi$ as is sometimes the custom. 

Equation \ref{eq:jfromp} is an intermediate step in the derivation of the Grad-Shafranov equation.\cite{grad_hydromagnetic_1958,solovev_plasma_1970} 

\subsection{Rigid rotor current condition, penetrated RMF}

The central assumption of current drive in a fully-penetrated Rotamak-FRC is that the current is due to all electrons rigidly rotating in synchrony with the applied RMF. The electron current is:
\begin{equation}
j_\phi = en_er\omega
\label{eq:jfromn}
\end{equation}
where $\omega$ is the angular rotational frequency of the applied RMF. In order to compute $P(\Psi)$, we will substitute Equation \ref{eq:jfromn} into Equation \ref{eq:jfromp}:
\begin{equation}
2\pi r\partial_\Psi P = en_er\omega
\label{eq:selfconsistent}
\end{equation}

If instead the current drive were due to ions rotating against stationary electrons, an ion momentum term would have to be added to the Grad Shafranov equation. These results still hold (with subscripts $e,i$ transposed) in the regime that $m_ir^2\omega^2\ll T_i$.

We will now discuss the relationship between density, temperature, and pressure:
\begin{equation}
n_iT_i+n_eT_e=nT=P
\label{eq:P}
\end{equation}
where 
\begin{equation}
n_e=n_i=n
\end{equation}
and
\begin{equation}
T=T_i+T_e
\label{eq:T}
\end{equation}
are valid when $Z=1$, the ion charge state is 1.

To proceed, we must make an assumption of the relative contributions of density and temperature to the pressure change. We use the common parametrization that density and temperature vary as a power law with the pressure whose exponents sum to 1:
\begin{equation}
n_e/n_0 = (P/P_0)^\mu
\label{eq:nfromp}
\end{equation}
\begin{equation}
T/T_0 = (P/P_0)^{1-\mu}
\label{eq:tofpanisothermal}
\end{equation}
where $\mu$ is a number between 0 and 1, and $P_0=n_0T_0$ are the pressure, density, and temperature at some arbitrary point. $\mu=0$ corresponds to the constant-density case, where variation in temperature is responsible for the variation in pressure. $\mu=1$ corresponds to the isothermal case, where variation in density is responsible for the variation in pressure. 

The behavior for $\mu=1$ must be treated differently from the behavior for $0\leq\mu< 1$.

\subsection{The case of $0\leq\mu<1$}

This case, encompassing all situations except the isothermal, was explored by R.G. Storer in 1982 and 1983.\cite{storer_pressure_1982,storer_compact_1983} The special case of $\mu=1/2$ was explored in detail by I.J. Donnelly \textit{et. al.} in 1987.\cite{donnelly_magnetohydrodynamic_1987}

Equations \ref{eq:selfconsistent} and \ref{eq:nfromp} have the solution:\cite{storer_pressure_1982}
\begin{equation}
P \propto (\Psi - \Psi_0)^\frac{1}{1-\mu}
\label{eq:pfrompsi}
\end{equation}
\begin{equation}
n_e \propto (\Psi - \Psi_0) ^\frac{\mu}{1-\mu}
\label{eq:nfrompsi}
\end{equation}
\begin{equation}
T = \frac{1-\mu}{2\pi} e\omega (\Psi - \Psi_0)
\label{eq:tfrompsi}
\end{equation}
where $\Psi_0$ is the value of the flux at the plasma-vacuum boundary, which may or not be the separatrix. $\Psi=0$ at the separatrix.

Several interesting features are apparent:

\paragraph*{Steep power law:}

Equation \ref{eq:pfrompsi} is what is substituted into Equation \ref{eq:jfromp} to find the Grad-Shafranov equilibrium. This is a power-law function, $P\propto\Psi^N$, $N=\frac{1}{1-\mu}\geq 1$. The least steep exponent is $N=1$, the Solov'ev function, which corresponds to $\mu=0$. This analysis indicates that the Solov'ev solution is only valid for rigid electron rotation when the density is constant, and only the temperature varies. If the density is allowed to vary at all, $N>1$ and the pressure flux function becomes more steep. We will find in Section \ref{sec:results} that a steep function of flux results in a pressure profile that is peaked at the magnetic axis.

\paragraph*{The specificity of $T$:}

$T$, as specified in Equation \ref{eq:tfrompsi}, does not have a multiplicative free parameter as $P,n$ do in Equations \ref{eq:pfrompsi} and \ref{eq:nfrompsi}. It is always linear to the flux, and the constant of proportionality is always $\frac{1-\mu}{2\pi} e\omega$. For $T$ to reach a large thermonuclear value, $\omega$ and $\Psi$ must be large enough. 

This analysis breaks down when $\mu=1$. This case is discussed in Section \ref{sec:isothermal}. The $\mu=1$ case is not $T=0$ as Equation \ref{eq:tfrompsi} would imply; rather it is $T$ constant. In order for the $\mu=1$ case to be the limiting case of Equation \ref{eq:tfrompsi} as $\mu\rightarrow1$, it must also be that $\Psi_0\rightarrow-\infty$.

Equation \ref{eq:tfrompsi} implies that $T$ can go to infinity as $\omega$ does the same. However, when $\omega\sim\Omega_e$, the cyclotron frequency of the electrons, then RMF no longer drives electrons and the assumption of rigid electron rotation is invalid. Equation \ref{eq:tfrompsi} should only be considered valid for $\omega<\Omega_e$.

\paragraph*{The free plasma boundary:}

The plasma-vacuum boundary is not necessarily the separatrix; there could be significant density in the open field line region outside the FRC. In fact this may be unavoidable, as transport of particles out of the FRC may fill this region.

\subsection{The isothermal case, $\mu=1$}
\label{sec:isothermal}

While Storer did not consider the case that the plasma could be isothermal, $\mu=1$, examination of this case actually precedes high-power RMF experiments. Christofilos and the Astron group modeled the Astron fusion reactor design using an isothermal rigid-rotor profile as early as the 1950s.\cite{christofilos_astron_1958,marder_bifurcation_1970} The well-known Rigid Rotor 1-D radial profile is implicitly isothermal.\cite{morse_equilibria_1969,armstrong_fieldreversed_1981,tuszewski_field_1988} 

These equilibria have been applied to non-RMF-driven FRC equilibria. The justification for using a rigid rotor model even when there is no RMF to drive to synchrony is often along the lines of Rostoker and Qerushi: ``the only drifted Maxwellians that satisfy the Vlasov equation for systems with cylindrical symmetry are rigid rotors."\cite{rostoker_equilibrium_2002} Because no RMF was assumed to drive the electrons at some angular velocity $\omega$, the value of $\omega$ was considered a free parameter, either assumed or used to fit to experimental measurements.

Several groups have written the Grad-Shafranov equation with a pressure flux function that corresponds to the isothermal case (exponential with flux), whether or not they explicitly recognized their equation as such.\cite{marder_bifurcation_1970,belova_numerical_2000,rostoker_equilibrium_2002,gota_separatrix_2003} 

Belova used the Grad-Shafranov equation with an isothermal pressure flux function to produce their starting FRC equilibria for analysis of stability.\cite{belova_numerical_2000} Gota used the Grad-Shafranov equation with an isothermal pressure flux function to fit to experimental data.\cite{gota_separatrix_2003} 

In the isothermal case, Equations \ref{eq:selfconsistent} and \ref{eq:nfromp} have the solution:
\begin{equation}
P \propto e^{\Psi/\Psi_c}
\label{eq:pexp}
\end{equation}
\begin{equation}
n_e \propto e^{\Psi/\Psi_c}
\label{eq:nexp}
\end{equation}
\begin{equation}
\Psi_c = \frac{2\pi T}{e\omega}
\label{eq:psic}
\end{equation}
where $\Psi_c$ is a characteristic flux determined by $T$ and $\omega$. $\Psi_c$ controls the steepness of the profiles. As $\Psi_c$ decreases, profiles become steeper. 

As a reminder, when temperature is not constant, Equations \ref{eq:pfrompsi} - \ref{eq:tfrompsi} hold rather than Equations \ref{eq:pexp} - \ref{eq:psic}. Several interesting features are apparent:

\paragraph*{Steep pressure function:}

Depending on the values of the factors in Equation \ref{eq:psic}, Equation \ref{eq:pexp} could be an extremely steep function of $\Psi$. As we will see in Section \ref{sec:results}, this translates into an extremely peaked pressure profile in space. 

Other researchers have noted that isothermal synchronous rotation can lead to a very peaked radial density profile $n(r)$.\cite{christofilos_astron_1958,rostoker_equilibrium_2002} However we will discuss in Section \ref{sec:results} that this steep profile has a tendency to be axially steep also, tightly peaked at the magnetic axis. 

\paragraph*{No plasma boundary:}
\label{sec:noboundary}

According to Equation \ref{eq:nexp}, there cannot be a plasma flux boundary outside of which the density is zero. There is always plasma outside the FRC separatrix, in the open field line region. This can be understood by examining the diamagnetic drift velocity where $n$ goes to zero as $T$ stays finite: it is locally infinite at this point and therefore cannot obey the rigid rotation criterion.

This lack of a boundary is visible even in the well-known Rigid Rotor 1-D radial profile, which exponentially decays to $n(r)\rightarrow0$ but never reaches it. 

Depending on the values of the factors in Equation \ref{eq:psic}, the drop-off of density outside the separatrix could be either steep, in which case plasma contact with the wall could be practically mitigated, or shallow, in which case wall contact is a large effect. We will explore this behavior in Section \ref{sec:results}

\paragraph*{Approximate density fall-off length:}

We may determine the density $e$-folding length from Equations \ref{eq:nexp} and \ref{eq:psic} evaluated at the magnetic field of the separatrix. The $e$-folding length ($n\propto e^{-r/L}$) is:
\begin{equation}
L = \frac{T}{rBe\omega}
\end{equation}

For an example case of $r=0.25$ m, $\omega=2\pi\times 10^6$ rad/s, $T=50$ keV, and $B=6$ T, $L=0.53$ m. One must be careful in this case to ensure that there is not too much density at the vacuum vessel wall. We will explore this behavior in more detail in Section \ref{sec:results}

\section{Likely values of $\mu$ in experiment and reactor}
\label{sec:likelymu}

The analysis presented in this paper assumes that the value of $\mu$ is known. Recall that $\mu$ is a measure of the relative peakedness of the density and temperature flux functions. At $\mu=0$, the density is constant and the temperature is peaked. At $\mu=1$, the temperature is constant and the density is peaked. In an experiment or a fusion reactor, several coupled processes will determine $\mu$. A few of these processes are: transport of particles and energy, wall interaction, and localized power deposition. 

Storer fits calculated equilibria to experimental data and obtains $N=1.6, \mu = 0.375$.\cite{storer_compact_1983} This was a small, cool ($T_e\approx 17$ eV) Rotamak which was wall-limited.\cite{euripides_rotamak_1997} Because the wall was cooled by edge contact, it is understandable that the pressure balance was determined more strongly by a temperature gradient than a density gradient ($\mu<0.5$).

In a fully-ionized, less collisional reactor-scale plasma, the situation will change. Transport of energy tends to be significantly faster than transport of particles.\cite{hinton_theory_1976} This will result in a density profile that is more peaked and a temperature profile that is more broad. This situation corresponds to $\mu>0.5$. In fact, if the recycling can be kept to a negligible level, the edge of the plasma may be at thermonuclear temperatures and the plasma may be effectively isothermal.\cite{stangeby_plasma_2000} Indeed, one philosophy of Tokamak design holds that a hot (thermonuclear temperature) edge is beneficial to fusion reactors.\cite{zakharov_li_2014}

Gota fits calculated equilibria to experimental data, evaluating the result for three assumed pressure flux functions.\cite{gota_separatrix_2003} The functions are the Solov'ev case ($P\propto\Psi,\mu=0$), the quadratic case ($P\propto\Psi^2,\mu=1/2$), and the isothermal case ($P\propto e^{\Psi/\Psi_0}),\mu=1$. They note that, for their experiment, the Grad-Shafranov equilibria using the three different profiles are ``almost the same."

\section{Applicability}
\label{sec:applicability}

The analysis presented in this paper is in the MHD regime. No kinetic effects are present. No ion flow is assumed. In an FRC-based compact fusion reactor, these effects may be important, as the ion thermal gryoradius is significant compared to the size of the plasma.

No analysis of the stability of these equilibria has been conducted.

The analysis presented in this paper assumes that the RMF has fully penetrated the plasma, the electrons are synchronously rotating with the RMF, and the ions are stationary. In actuality several effects may make this inapplicable to experiments or reactors.

RMF may not fully penetrate the plasma if the RMF magnitude is too weak, the plasma is too collisional, or the plasma radius is too large.\cite{hugrass_production_1979,hugrass_numerical_1981,jones_steady-state_1981,jones_review_1999,milroy_numerical_1999,hoffman_principal_2006} RMF will penetrate only to a certain radius. Hugrass uses this penetration length:\cite{hugrass_production_1979,hugrass_numerical_1981}
\begin{equation}
\delta_{RMF}=\frac{\omega_{ce,RMF}}{\nu_{e,i}}\sqrt{\frac{\eta}{\omega_{RMF} \mu_0}}
\label{eq:penetration}
\end{equation}
where $\delta_{RMF}$ is the penetration depth of the RMF field, $\omega_{ce,RMF}$ is the electron gyrofrequency in the RMF field, $\nu_{e,i}$ is the electron-ion collision time, $\eta$ is the resistivity of the plasma, $\omega_{RMF}$ is the angular frequency of the RMF, and $\mu_0$ is the magnetic permeability of free space.

Using the Spitzer resistivity for $\eta$, we find the following dependency:
\begin{equation}
\delta_{RMF}=\omega_{ce,RMF} \sqrt{\frac{3\alpha_S}{2^{9/2}\pi^{3/2}}\frac{1}{\omega_{RMF}} \frac{(T_e/m_e c^2)^{3/2}}{Z n_e^2 r_e^3 c  \ln{\Lambda}}}\propto n^{-1}
\label{eq:better-units}
\end{equation}
where $\alpha_S\approx 0.51$ is the Spitzer correction to the DC resistivity, $m_e/c^2\approx511\times10^{3}$ eV is the electron rest energy, $r_e\approx2.82\times10^{-13}$ cm is the classical electron radius, and $c\approx3.00\times10^{10}$ cm/s is the speed of light in a vacuum.

Using example parameters of $T_e=50$ keV, $n_e = 4\times10^{14}$ /cc, $\omega_{RMF} =2\pi\times10^6$ rad/s, $B_{RMF}=16$ Gauss, we find that $\delta_{RMF}$ = 50 cm.

The Grad-Shafranov equation does not include the anisotropic pressure effects which give rise to mirror axial confinement. This analysis does not take into account these mirror confining effects and so the pressure profile outside the separatrix may be different than those determined by the Grad-Shafranov equation. However, the argument that $n\rightarrow0$ at constant $T$ is incompatible with rigid rotation is still valid.

\section{The solver: Iteratively determined Grad-Shafranov Equilibria}
\label{sec:solver}

Two equations were Picard iterated to determine the self-consistent Grad-Shafranov equilibrium. One of them was Equation \ref{eq:jfromp}, reproduced here with more explicit dependences:
\begin{equation}
j_\phi(r,z) = 2\pi r \partial_\Psi P (\Psi(r,z))
\label{eq:jiter}
\end{equation}

The other is $\Psi$ generated from the resulting $j_\phi$, as determined from the elliptic integral Green's function of Ampere's Law for flux in cylindrical coordinates:
\begin{equation}
\Psi(r,z) = \int dr^\prime \int dz^\prime j_\phi(r^\prime,z^\prime) G(r,z,r^\prime,z^\prime) + \Psi_v + \Psi_{FC}
\label{eq:psiiter}
\end{equation}
where $\Psi_v$ is the vacuum flux and $\Psi_{FC}$ is the flux from flux-conserving current loops, if any.

An initial guess for $\Psi$ was determined heuristically. Equations \ref{eq:jiter} and \ref{eq:psiiter} were successively applied to the existing $j_\phi$ and $\Psi$ guesses until the variation was smaller than a tolerance. In this manner a self-consistent equilibrium was computed. 

Useful Grad-Shafranov solvers must include the possibility that some axial field coils conserve magnetic flux. On a short timescale, all electrically conductive loops such as the vacuum vessel wall will conserve flux. On a long timescale, any superconducting coils operating in a persistent mode will conserve flux. 

For computations including flux conserving current loops, the flux conserver current $I_{FC}$ was determined using the equation
\begin{equation}
\vec{I}_{FC} = M^{-1} \vec{\Psi}_{P,FC}
\end{equation}
where $\vec{I}_{FC}$ is the list of flux conserver currents, $\vec{\Psi}_{P,FC}$ is the list of plasma fluxes computed from $j_\phi$ evaluated at the flux conserver locations, and $M$ is the matrix of mutual- and self-inductances between the flux conserving loops.

The function $\partial_\Psi P (\Psi)$ in Equation \ref{eq:jiter} comes from either Equation \ref{eq:pfrompsi} or \ref{eq:pexp} (if isothermal). Both of these equations have a free multiplicative factor. This factor can be assumed, or can be used to satisfy a useful constraint, such as a location that lies upon the separatrix or the maximum value of the flux. If this approach is to be used, the value of the free factor is set every iteration after Equation \ref{eq:jiter} is applied, by enforcing the constraint. Other constraints might be: Diamagnetic loop measurement constrained to be a specific value, line-averaged density constrained to be a specific value, maximum flux constrained to be a specific value, etc.

Any configuration of axial field coils (producing $\Psi_v$) and flux-conserving loops (producing $\Psi_FC$) may be used. Each corresponds to a different experiment or reactor. For generality, the results given in this paper are for constant vacuum magnetic field, $\Psi_v\propto r^2$, and unless otherwise stated there were no flux conserving loops. 

\section{Results}
\label{sec:results}

\subsection{The case of $0\leq\mu<1$}
\label{sec:anisothermalresults}

Several equilibria were found for various $N=\frac{1}{1-\mu}$ values, corresponding to various dependencies of the density and temperature on the pressure. $N=1$ corresponds to the constant-density, varying-temperature case, and the Solov'ev solution is recovered. As $N$ increases, the density profile becomes more and more peaked compared to the temperature profile.

The equilibria were computed assuming a uniform vacuum field of 5 Tesla and an RMF frequency of $2\pi\times0.5\times10^{6}$ rad/s. The separatrix radius was constrained to be 20 cm. 

\paragraph*{Increasing oblateness:}

\begin{figure}
\includegraphics[width=0.5\textwidth]{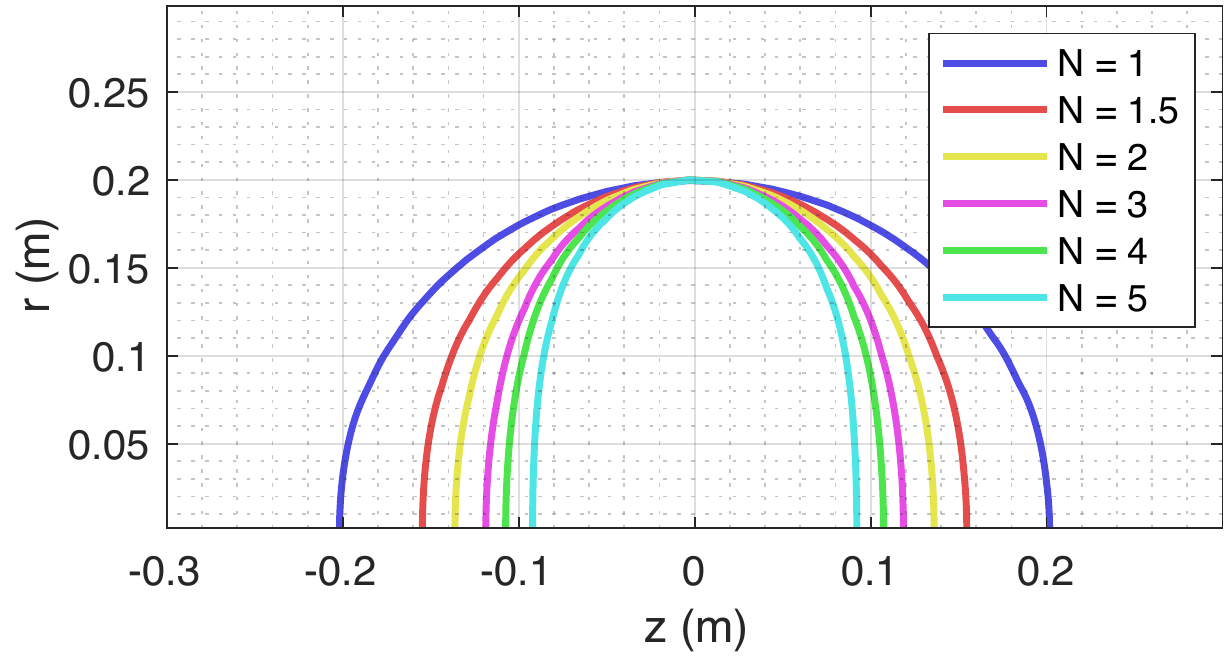}
\caption{\label{fig:exponentseparatrices} FRC separatrices for various values of $N$, where $P\propto\Psi^N$. Different values of $N$ correspond to different relative peakednesses of density and temperature.}
\end{figure}

Figure \ref{fig:exponentseparatrices} shows the separatrices of FRCs calculated with several values of $N$. $N=1$ corresponds to the Solov'ev case, and a spherical Hill's Vortex is recovered. As $N$ increases, the separatrix becomes more and more oblate. 

\begin{figure}
\includegraphics[width=0.5\textwidth]{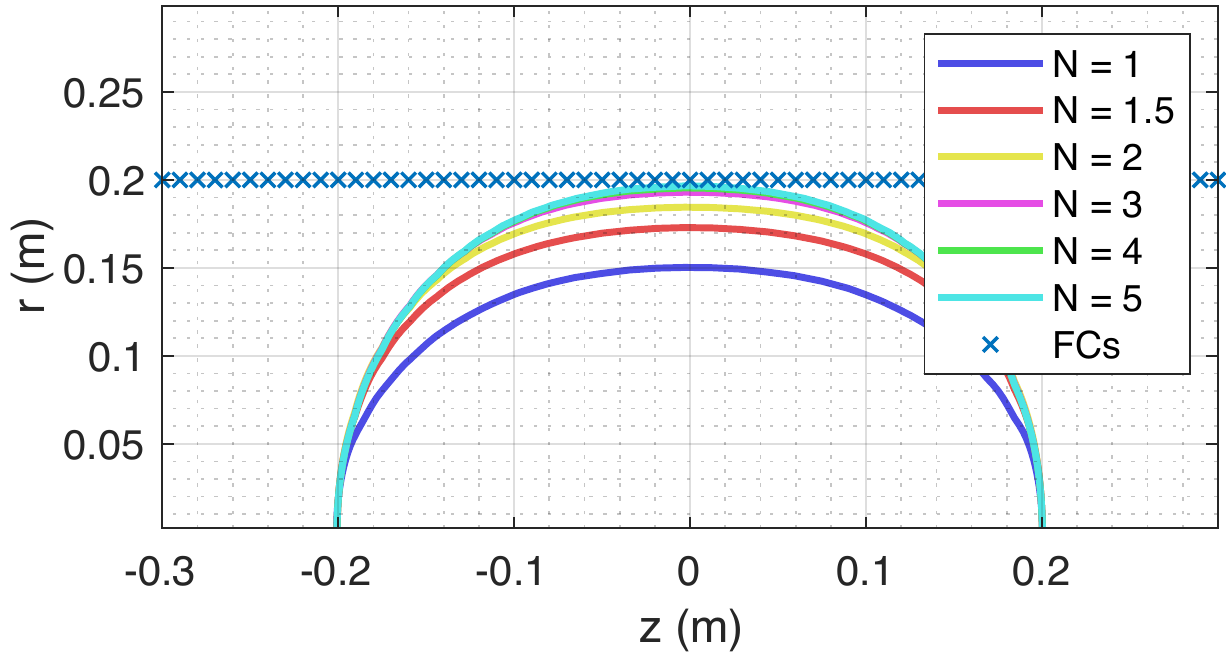}
\caption{\label{fig:exponentFCseparatrices} FRC separatrices for various values of $N$. A barrier of flux-conserving loops has been placed at $r=20$ cm. These flux conservers (FCs) are able to counteract the tendency for the FRC to become oblate.}
\end{figure}

This oblateness can be mitigated with the use of flux conserving current elements in close proximity to the FRC. Another set of solutions is depicted in Figure \ref{fig:exponentFCseparatrices}. The difference is that a cylindrical shell of closely spaced, flux conserving loops was placed around the plasma, constraining its radial growth. For these solutions, the X-point was constrained to lie at 20 cm. As can be seen in that figure, flux conservers are able to keep the FRC prolate. By tailoring the placement of axial field coils and flux conserving loops, it is possible to control the shape of the plasma separatrix.

\begin{figure}
\includegraphics[width=0.5\textwidth]{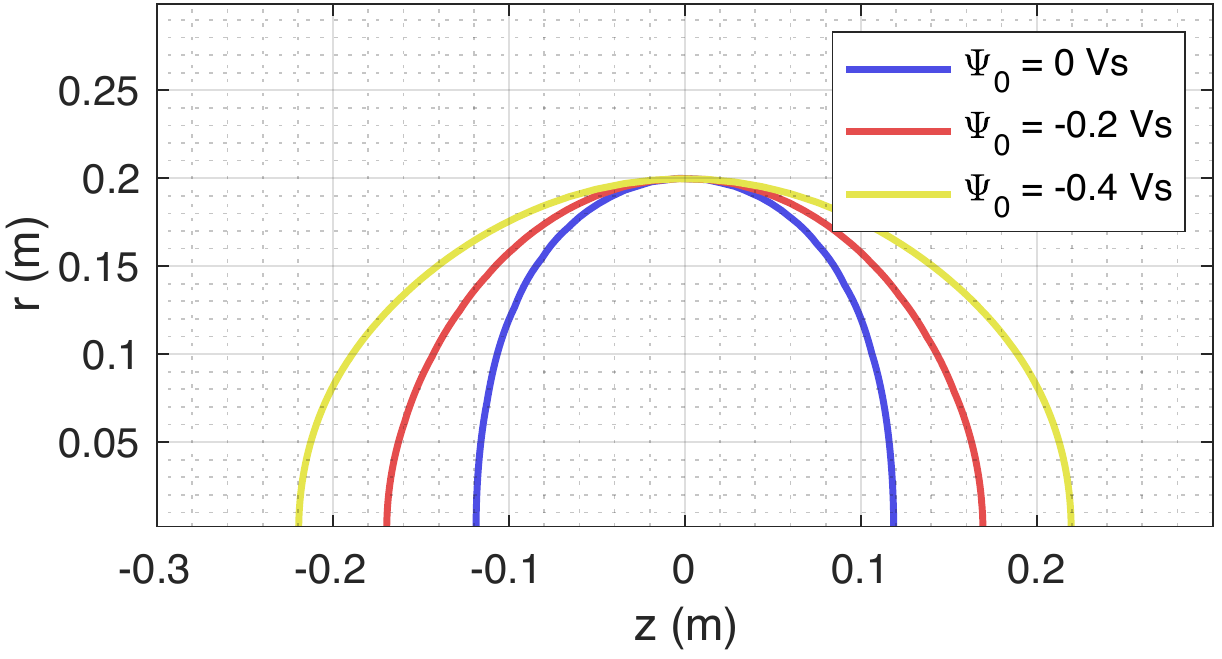}
\caption{\label{fig:exponentfluxlimseparatrices} FRC separatrices for $N=3$. Various values of the limiting flux $\Psi_0$ are used. Outside this flux, $n,T=0$. As plasma is allowed to exist outside the separatrix ($\Psi_0$ becomes negative), the separatrix becomes less oblate.}
\end{figure}

\begin{figure}
\includegraphics[width=0.5\textwidth]{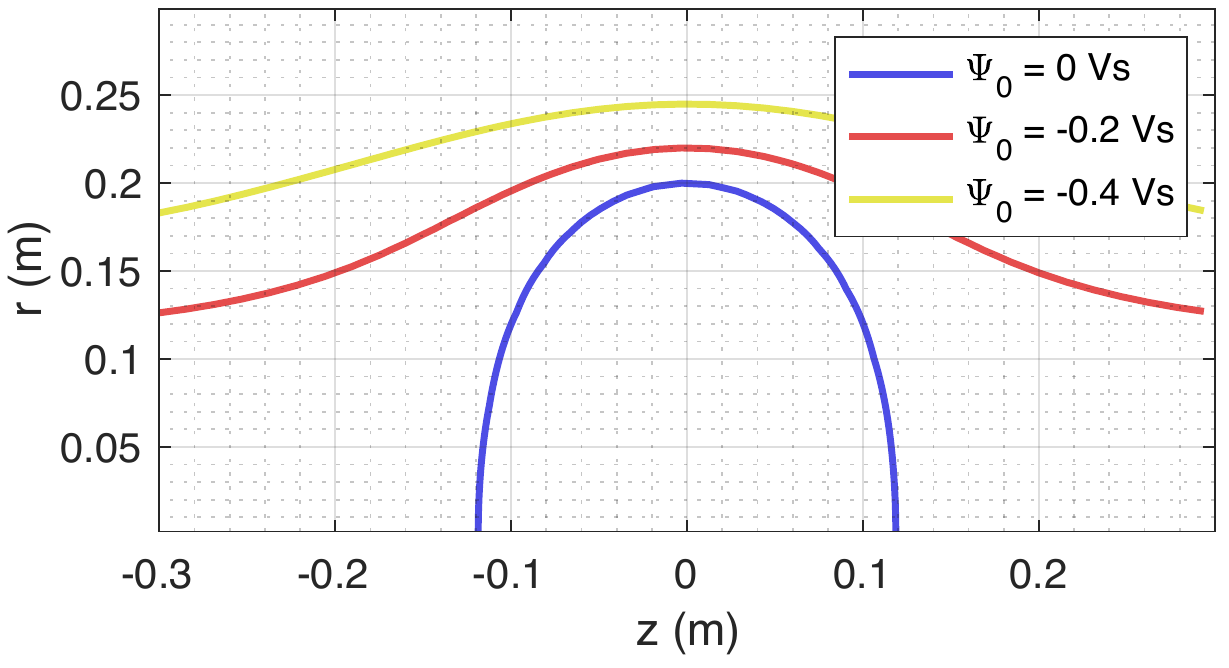}
\caption{\label{fig:exponentfluxlimfluxlimits} The same equilibria as Figure \ref{fig:exponentfluxlimseparatrices}. The spatial location at which $\Psi_0$ is reached for various values of $\Psi_0$. Outside these contours, $n,T=0$. As $\Psi_0$ becomes more negative, the plasma persists farther outward of the separatrix.}
\end{figure}

Yet another set of solutions is depicted in Figure \ref{fig:exponentfluxlimseparatrices}. This set of solutions keeps $N=3$ and varies the $\Psi_0$ parameter in Equation \ref{eq:pfrompsi}, the flux limit outside of which the density and temperature are zero. As density and temperature is allowed to exist outside the separatrix ($\Psi_0$ becomes negative), the separatrix becomes less oblate and more prolate. However, there is significant density outside the FRC, where plasma is less well confined and it can hit the walls or flow to a divertor or end cell. The flux limit, $\Psi_0$, is shown in Figure \ref{fig:exponentfluxlimfluxlimits}. As $\Psi_0$ becomes more negative, more of the plasma is in the open field line region and approaches the wall of the vacuum vessel.

It may be that transport requires $\Psi_0<0$ in experiments and reactors. This would mean that there is always some amount of plasma outside the separatrix, in the open field line region. Confinement is poorer in the open field line region. It is mirror confinement rather than cross-field confinement, causing axial losses. Some implications of this are discussed briefly in Section \ref{sec:conclusion}.

\paragraph*{Increasingly peaked density:}

\begin{figure}
\includegraphics[width=0.5\textwidth]{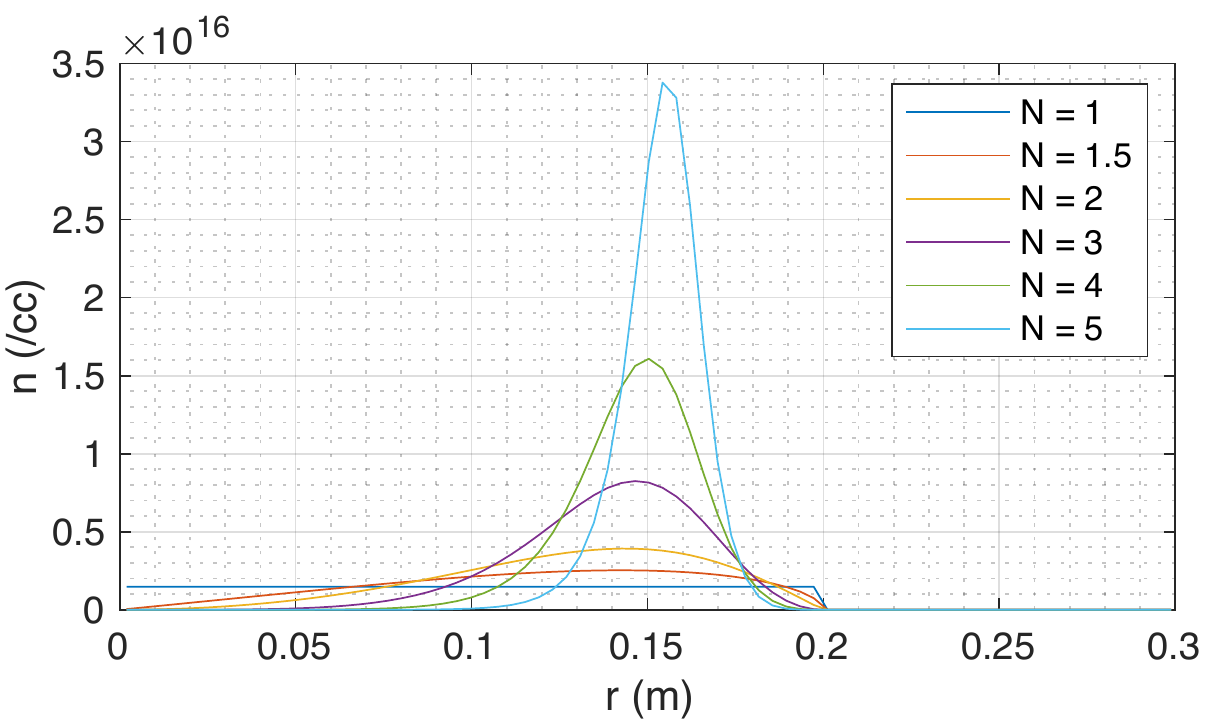}
\caption{\label{fig:exponentdensities} The same equilibria as Figure \ref{fig:exponentseparatrices}. Radial profiles of density $n(r)$ at the $z=0$ plane for various values of $N$. As $N$ increases, the density profile becomes more peaked and the maximum density increases. We have assumed that $n_e=n_i=n$.}
\end{figure}

Figure \ref{fig:exponentdensities} shows the radial profile of the density $n(r)$ at $z=0$. For the $N=1$ case, the density is constant as was assumed. As $N$ increases, the maximum density grows larger and the density profile becomes more peaked and narrow. 

As discussed in Section \ref{sec:applicability}, at some point the increasing density will cause imperfect penetration of the RMF, saturating the effect and limiting the density peakedness.

The fact that the density is more peaked is not of itself useful. We will compute its effect on the volume-averaged pressure and a quantity relevant to fusion power density in the next subsections. Surprisingly, locally the density profile is so peaked that $\beta\gg1$ over a small volume. 

\paragraph*{Increasingly peaked temperature:}

\begin{figure}
\includegraphics[width=0.5\textwidth]{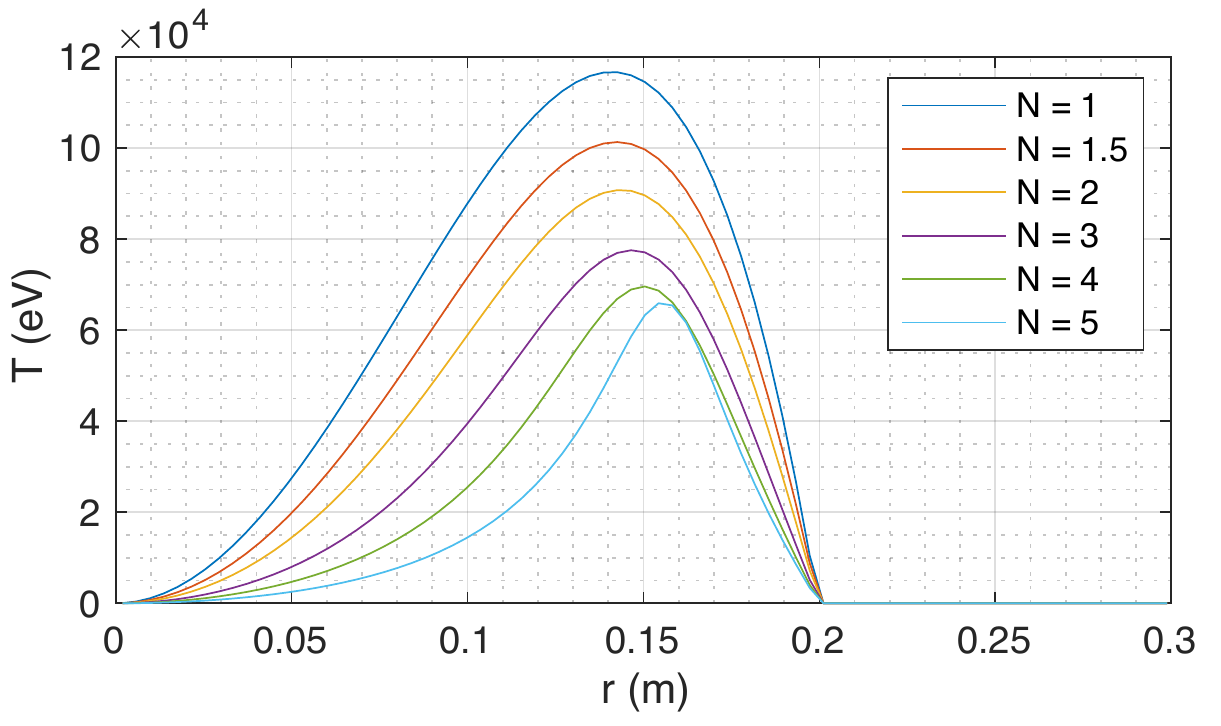}
\caption{\label{fig:exponenttemperatures} The same equilibria as Figure \ref{fig:exponentseparatrices}. Radial profiles of temperature $T(r)$ at the $z=0$ plane for various values of $N$. As $N$ increases, the temperature profile becomes more peaked and the maximum temperature decreases. These high temperatures, $T>100$ keV, are relevant to advanced fuels such as D+$^3$He.}
\end{figure}

Figure \ref{fig:exponenttemperatures} shows the radial profile of the temperature $T(r)$ at $z=0$. Recall that $T=T_e+T_i$ as defined in Equation \ref{eq:T}. For the $N=1$ case, the temperature is proportional to the flux $\Psi$. As $N>1$, $T(r)$ becomes more peaked, though less so than $n(r)$.

This is a surprising result. One might instead expect $T(r)$ to become less peaked as $N$ increases, as the dependence of temperature on pressure $T(P)$ becomes less steep as per Equation \ref{eq:tofpanisothermal}. However, as $N$ increases the pressure flux function $P(\Psi)$ becomes steeper as per Equation \ref{eq:pfrompsi}. The net effect is that the pressure profile $P(r)$ as determined via Picard iteration becomes steeper faster than the temperature dependence on pressure $T(P)$ becomes shallow, and the net effect is that the temperature profile $T(r)$ becomes more steep.

\paragraph*{Decreasing volume-averaged plasma pressure:}

\begin{figure}
\includegraphics[width=0.5\textwidth]{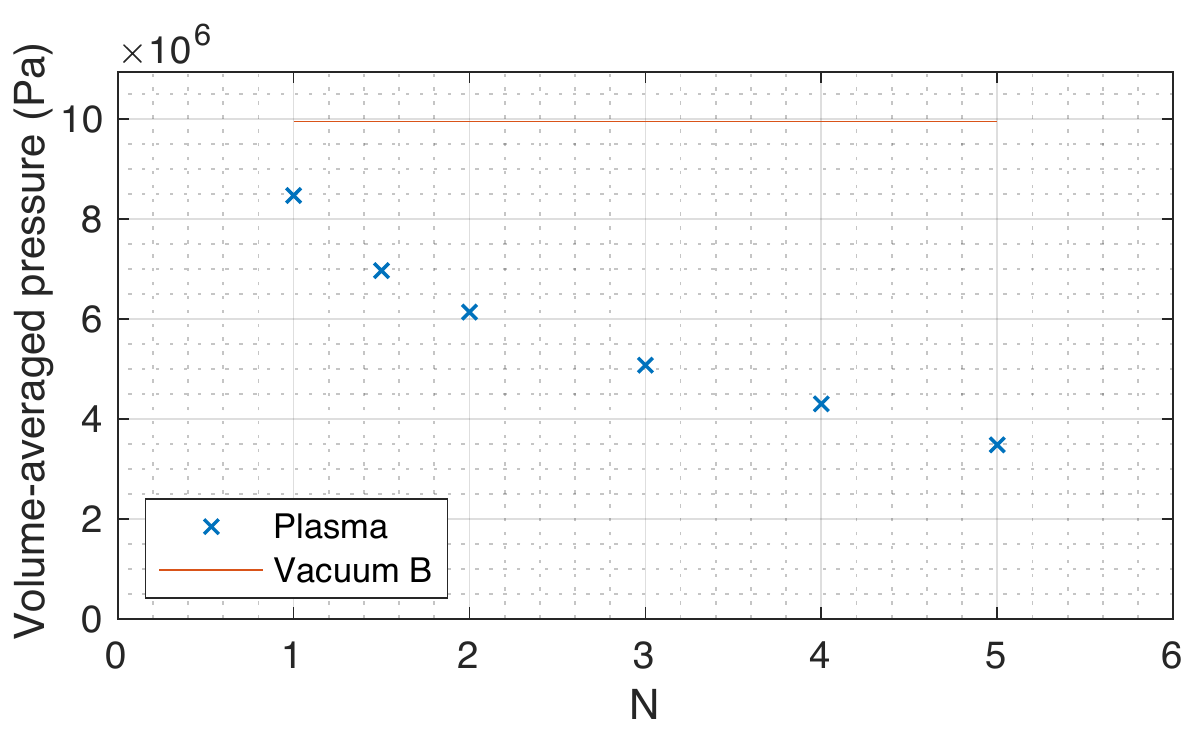}
\caption{\label{fig:exponentpressures} The same equilibria as Figure \ref{fig:exponentseparatrices}. These are the volume-averaged plasma pressure for various values of $N$. The volume was taken to be a cylinder with radius $r=20$ cm, the separatrix radius, and $L=40$ cm long. Recall there is no plasma outside the separatrix in these equilibria. The volume-averaged pressure decreases with increasing $N$.}
\end{figure}

Figure \ref{fig:exponentpressures} shows the plasma pressure $P$, averaged over a cylinder with the radius of the separatrix and the half-length of the radius of the separatrix (20 cm). While the maximum density clearly increases, as can be seen in Figure \ref{fig:exponentdensities}, it is squeezed into an ever smaller volume, and so the volume-averaged pressure decreases. Consequently the FRC has a lower volume-averaged pressure ratio, $\langle \beta \rangle$, at higher $N$. This may at first seem deleterious to a fusion reactor. However, the fusion power density is not proportional to plasma pressure $P$; rather it is proportional to $n^2$ with a highly nonlinear function of $T$.

\paragraph*{Increasing volume-averaged square pressure:}

\begin{figure}
\includegraphics[width=0.5\textwidth]{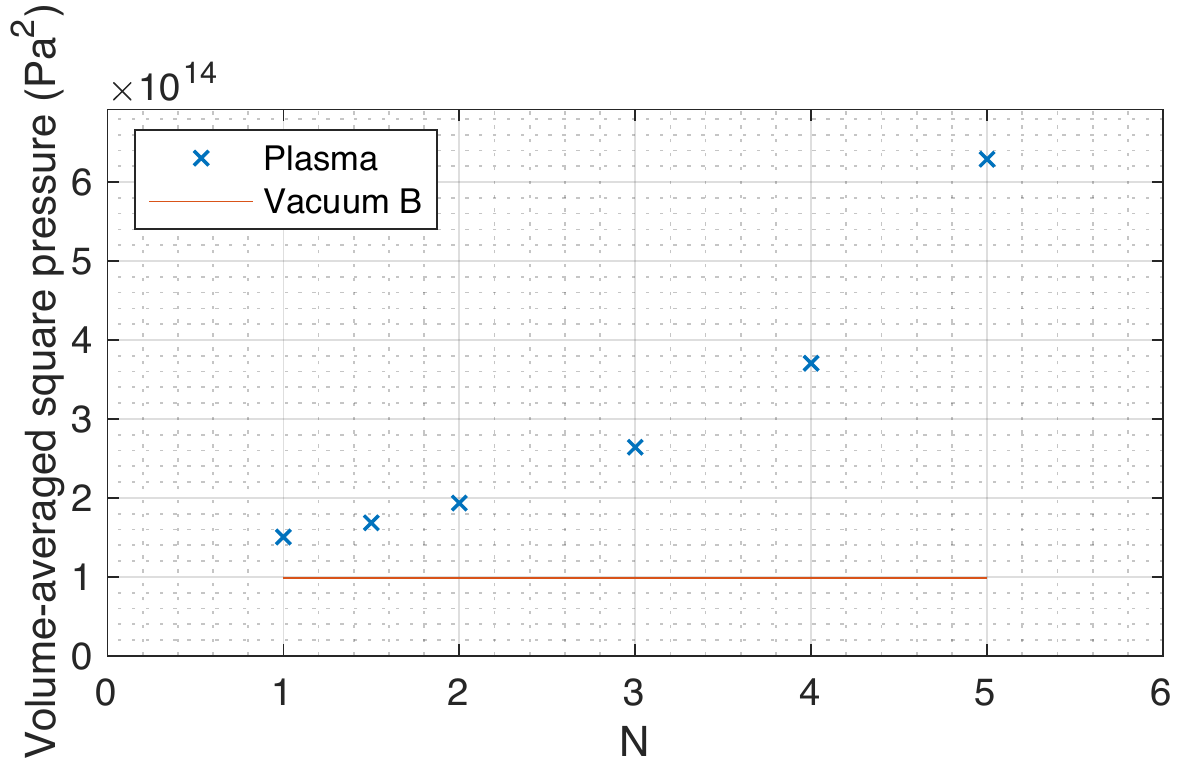}
\caption{\label{fig:exponentpressuressq} The same equilibria as Figure \ref{fig:exponentseparatrices}. These are the volume-averaged square plasma pressure for various values of $N$. The volume was taken to be a cylinder with radius $r=20$ cm, the separatrix radius, and $L=40$ cm long. The volume-averaged square pressure increases with increasing $N$. Recall there is no plasma outside the separatrix in these equilibria. The volume-averaged square pressure is a good approximation for the fusion power when the temperature profile is less peaked than the density profile.}
\end{figure}

Figure \ref{fig:exponentpressuressq} shows the square plasma pressure, $\langle n^2T^2 \rangle$, averaged over a cylinder with the radius of the separatrix and the half-length of the radius of the separatrix (20 cm). It is an increasing function of $N$. In the balance between increasing density and decreasing volume, the increasing density wins out and the quantity increases. 

These values of $\langle n^2T^2\rangle$ were generated using $\omega = 0.5 \times 2\pi \times 10^6$ rad/s. Flux field $\Psi$ and pressure $P$ were produced via Picard iteration. Temperature is specified per Equation \ref{eq:tfrompsi}, which then specifies density $n$ via the pressure relationship, Equation \ref{eq:P}. Recall that $\omega$ is essentially a free parameter. Thus, $\omega$ could be scaled so that $T_{max}$, the maximum temperature, were constant in $N$. 

At high $N$, density is much more peaked than $T$. We can therefore approximate $T\approx T_{max}$ as constant over the region of high $n^2$. Applying the procedure in the preceding paragraph ($T_{max}$ constant), this quantity $\langle n^2T^2 \rangle \approx T_{max}^2\langle n^2\rangle \propto n^2$ therefore approximates the fusion power density. 

The results in Figure \ref{fig:exponentpressuressq} indicate that the fusion power output from a Rotamak-FRC whose temperature is more constant than its density ($\mu>0.5$) can be higher than the power output from an equivalent volume of plasma with $\beta=1$ (plasma pressure over vacuum field). The concentration of density into a peaked structure is responsible for this result. 

\paragraph*{Summary table:}

A summary of these results is shown in Table \ref{tab:summary}. They are compared to a case called $N=0$ but is simply the result of a point-plasma model where $P=nT$ is determined from $\beta=1$ in the 5 Tesla vacuum field. For more peaked density than temperature, which is likely for reactor-scale plasmas, the fusion power output can be much higher than the equivalent volume of $\beta=1$ plasma. $\beta$ is calculated with respect to the vacuum field.

\begin{table}
\caption{\label{tab:summary} Summary of the results of Figures \ref{fig:exponentpressures} and \ref{fig:exponentpressuressq}. The $N=0$ case is the result of a point-plasma model, where the pressure was calculated from $\beta=1$ in the 5 Tesla vacuum field. The $N=1$ case is the commonly assumed Solov'ev solution, the Hill's Vortex. F is $\langle\beta^2\rangle$, which approximates the enhancement to the fusion power as a result of density peakedness. $N$ is a measure of the relative peakedness of the density and temperature; $N>1$ is likely in reactor-scale experiments as discussed in Section \ref{sec:likelymu}.}
\begin{ruledtabular}
\begin{tabular}{llll}
$N$&$\langle P\rangle$&$\sqrt{\langle P^2\rangle}$ & F\footnote{Approximate fusion power enhancement factor, $\langle \beta^2\rangle$}\\
\hline
0\footnote{Point-plasma model. Uniform pressure, $\beta=1$ in 5 T vacuum field} & 9.95 MPa & 9.95 MPa & 1 \\
1\footnote{Solov'ev solution, Hill's Vortex} & 8.47 MPa & 12.3 MPa & 1.52 \\
1.5 & 6.96 MPa & 13.0 MPa & 1.70 \\
2 & 6.13 MPa & 13.9 MPa & 1.96 \\
3 & 5.07 MPa & 16.3 MPa & 2.67 \\
4 & 4.30 MPa & 19.2 MPa & 3.74 \\ 
5 & 3.48 MPa & 25.1 MPa & 6.36 \\
\end{tabular}
\end{ruledtabular}
\end{table}

\subsection{The isothermal case, $\mu=1$}

In this section we will discuss the special case of an isothermal plasma, $\mu=1,N=\infty$.

The isothermal equilibria are characterized by the parameter $\Psi_c$ in Equation \ref{eq:pexp}. Several equilibria were found for various $\Psi_c = \frac{2\pi T}{e\omega}$ values, corresponding to various temperatures and RMF frequencies. 

The equilibria were computed assuming a uniform vacuum field of 5 Tesla and a temperature of 50 keV. The separatrix radius was constrained to be 20 cm. The values of $\Psi_c = [0.115, 0.130, 0.200]$ Vs correspond to RMF angular frequencies of $\omega = [2.73\times10^6, 2.42\times10^6, 1.57\times10^6]$ rad/s respectively.

Values of $\Psi$ are given in Vs, or Volt-Seconds. This is equivalent to Tesla-meter-squared.

Values of $\Psi_c$ less than 0.115 Vs produced numerical problems, as the discretization of the grid (8 mm) was too large, so these equilibria could not be computed accurately. As discussed in Section \ref{sec:applicability}, at some point the increasing density will cause imperfect penetration of the RMF, saturating the effect and limiting the density peakedness.

\paragraph*{Prolate and oblate separatrix:}

\begin{figure}
\includegraphics[width=0.5\textwidth]{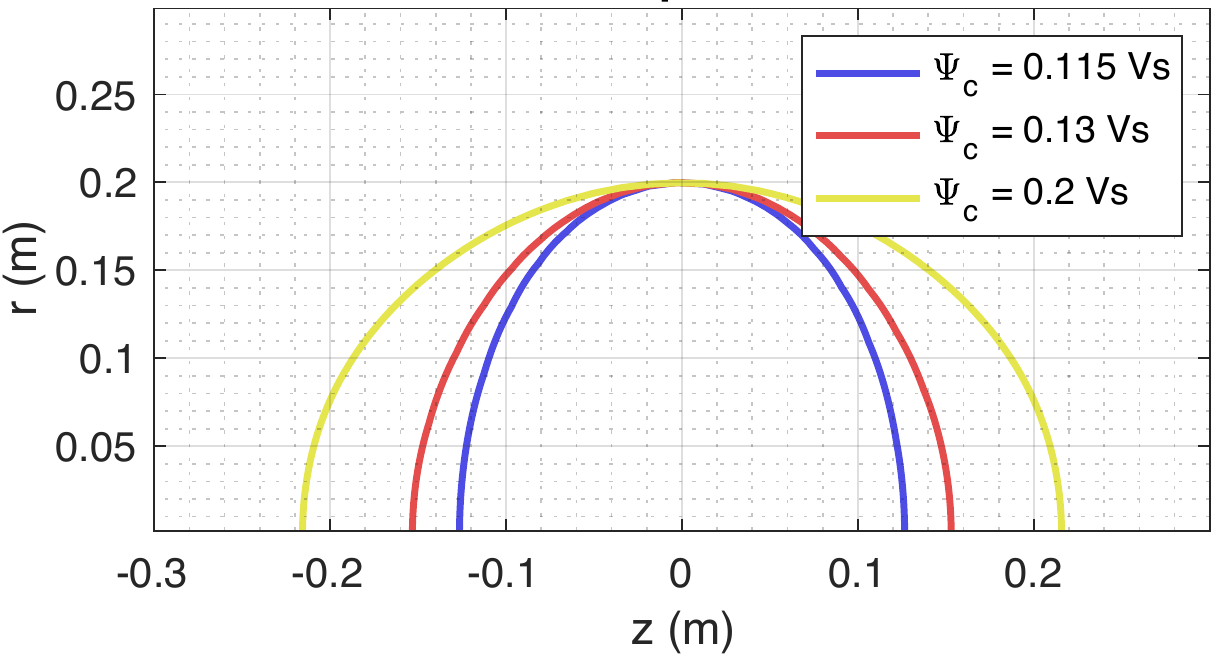}
\caption{\label{fig:isothermalseparatrices} FRC separatrices for various values of $\Psi_c$, where $P\propto e^{\Psi/\Psi_c}$. Different values of $\Psi_c$ correspond to different values of $T,\omega$. As $\Psi_c$ decreases ($T$ decreases or $\omega$ increases), the FRC becomes more oblate.}
\end{figure}

Figure \ref{fig:isothermalseparatrices} shows the separatrices for various values of $\Psi_c$. The FRC can be either naturally oblate or naturally prolate, depending on the value of $\Psi_c$. As with the non-isothermal case, the shape of the FRC can also be manipulated with flux conserving or current-carrying coils (not shown). 

\paragraph*{Peakedness of density:}

\begin{figure}
\includegraphics[width=0.5\textwidth]{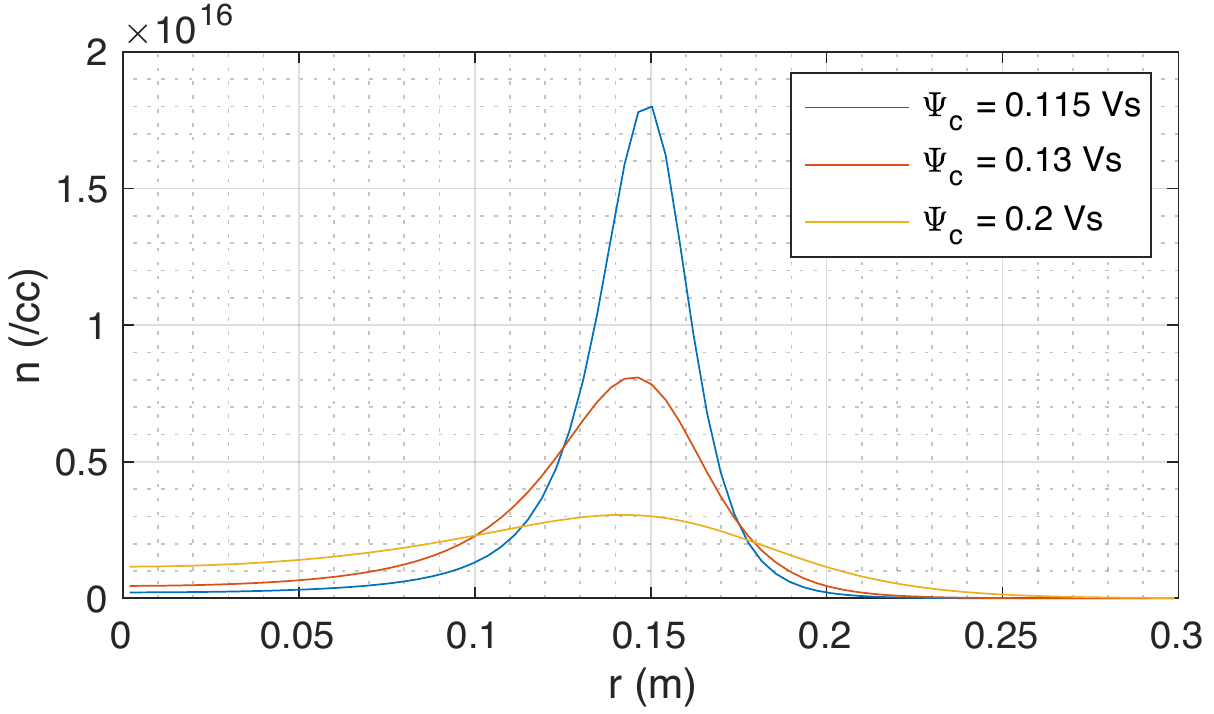}
\caption{\label{fig:isothermaldensities}The same equilibria as Figure \ref{fig:isothermalseparatrices}. Radial profiles of density $n(r)$ at the $z=0$ plane for various values of $\Psi_c$. As $\Psi_c$ decreases, the density profile becomes more peaked and the maximum density increases. Locally, this highly peaked density can cause small volumes of $\beta\gg 1$. We have assumed that $n_e=n_i=n$.}
\end{figure}

Figure \ref{fig:isothermaldensities} shows the radial density profiles $n(r)$ along the $z=0$ line for various values of $\Psi_c$. A small $\Psi_c$ corresponds to a high maximum density and a peaked spatial profile. A large $\Psi_c$ corresponds to a low maximum density and a broad spatial profile. 

As discussed in Section \ref{sec:applicability}, at some point the increasing density will cause imperfect penetration of the RMF, saturating the effect and limiting the density peakedness.

\begin{figure}
\includegraphics[width=0.5\textwidth]{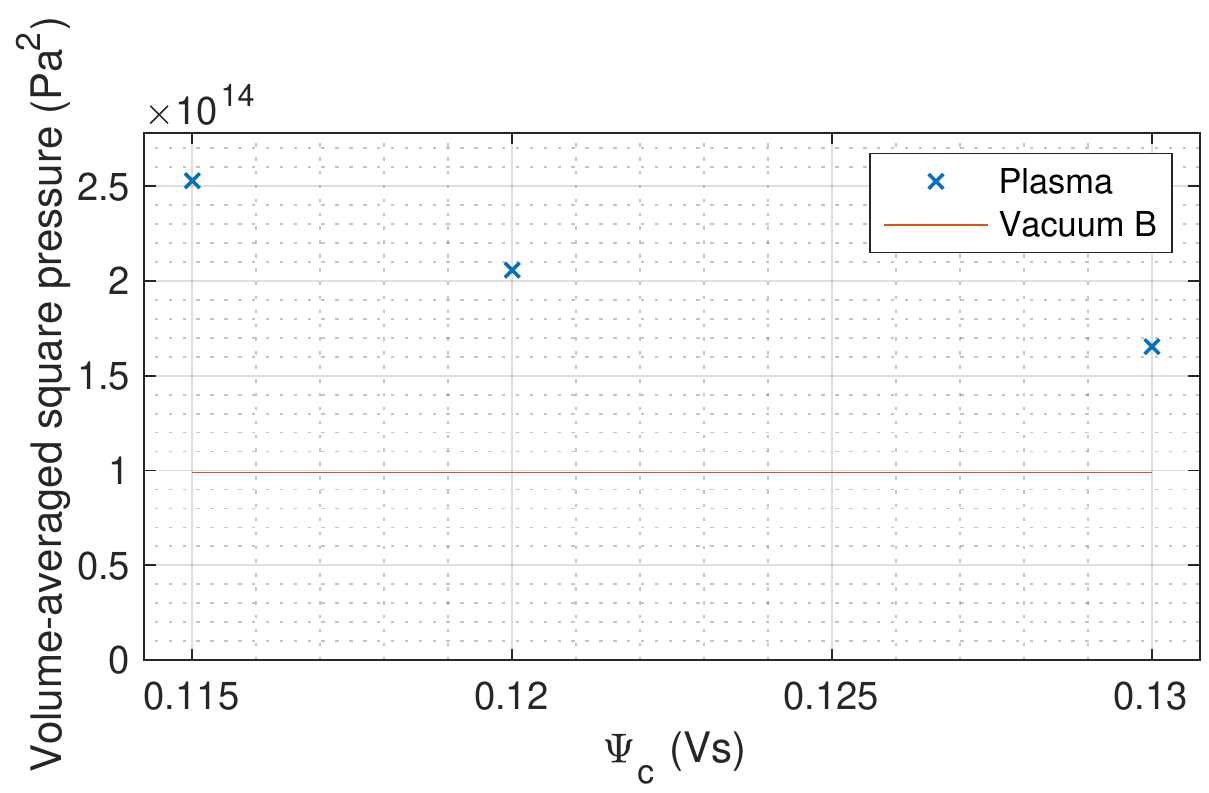}
\caption{\label{fig:isothermalpressuressq} A set of isothermal equilibria with differing $\Psi_c$. These are the volume-averaged square plasma pressure. The volume was taken to be a cylinder with radius $r=20$ cm, the separatrix radius, and $L=40$ cm long. The volume-averaged square pressure decreases with increasing $\Psi_c$. The volume-averaged square pressure is proportional to the fusion power when the temperature is constant.}
\end{figure}

As discussed in Section \ref{sec:anisothermalresults}, this density peakedness also implies a higher fusion rate. As $\Psi_c$ decreases ($\omega$ increases), the density profile becomes more and more peaked, decreasing $\langle\beta\rangle$ but increasing $\langle\beta^2\rangle$, which corresponds to fusion reaction rate. This is shown in Figure \ref{fig:isothermalpressuressq}. At $\Psi_c=0.115$ Vs, the numerical stability limit for the resolution used (8 mm), the fusion rate is enhanced a factor of 2.5 over a $\beta=1$ uniform plasma volume. As with the non-isothermal case, more peaked implies more fusion. 

\paragraph*{Density fall-off outside the separatrix:}

\begin{figure}
\includegraphics[width=0.5\textwidth]{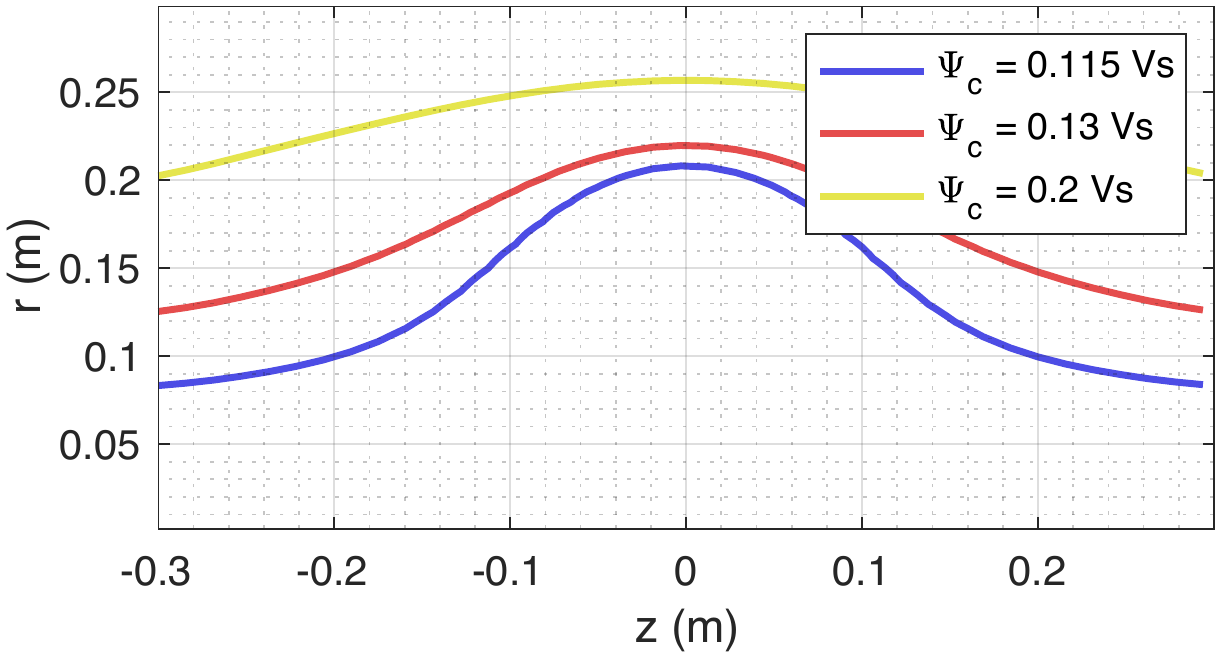}
\caption{\label{fig:isothermaldensitythresholds} The same equilibria as Figure \ref{fig:isothermalseparatrices}. These are the contours at which $n_e=10^{14}$ /cc. This indicates that the high density region extends farther toward the wall of the vacuum vessel when $\Psi_c$ increases. }
\end{figure}

Recall from Section \ref{sec:noboundary} that $n(r)\rightarrow0$ is not compatible with isothermal rigid rotation. Thus, it can be difficult to keep the plasma away from the walls. Figure \ref{fig:isothermaldensitythresholds} shows the contours in space where the density falls below $10^{14}$ /cc. For small $\Psi_c$, the exponential fall-off is sufficient to reduce the density to below $10^{14}$ /cc in a short distance from the separatrix. As $\Psi_c$ grows larger, the density profile broadens and the walls must be placed farther and farther away.

Confinement is poorer in the open field line region. It is mirror confinement rather than cross-field confinement, causing axial losses. Some implications of this are discussed briefly in Section \ref{sec:conclusion}.

Even with $\Psi_c=0.115$ Vs, the field lines of the magnetic axis have density higher than $10^{14}$ /cc. These field lines intersect with the vessel wall in a divertor or end cell. There, the power flux will be high. 

\section{Discussion and conclusion}
\label{sec:conclusion}

We have determined that fully co-rotating RMF-driven Rotamak-FRC experiments and reactors could beneficially have much more sharply peaked density profiles than a Hill's Vortex, the Solov'ev solution. This is because the temperature profile is likely to be broader than the density profile, which together with rigidly rotating electrons implies a sharper functional form of the Grad-Shafranov pressure flux function. This entails both benefits and challenges to the creation of a compact fusion reactor. 

One benefit is that the peakedness of the density profile enables a substantial increase in the total fusion power of the reactor, even while decreasing the $\langle\beta\rangle$. This increase may be $\sim5-10\times$ depending on the relative broadness of the temperature and density. 

One challenge is that the separatrix of the FRC tends to be naturally oblate. For a variety of reasons, this may not be desired. This oblateness can be mitigated with separatrix shaping from current-carrying or flux-conserving coils around the FRC. It may also be mitigated by allowing there to be significant plasma density outside the FRC, in the open field line region.

Another challenge is that density outside the FRC may be unavoidable. In the isothermal case, there must be density outside the separatrix, but wall contact can be reduced with appropriate choice of RMF frequency. In the non-isothermal case, density outside the separatrix is not ruled out and may be populated by transport out of the FRC. 

A limiter or high the open field line region's high axial losses could enforce a step-like jump in density. In this case, the electron diamagnetic velocity becomes infinite at the jump, and so is faster than the local rigid-rotor velocity. If the RMF acts to slow down these fast-moving electrons as efficiently as it speeds up slow-moving electrons to the RMF velocity, then the RMF will act to broaden the density profile at the jump rather than steepen it. This process may be slow and negligible compared to transport processes.

This model allows RMF to produce arbitrarily peaked density profiles at high RMF frequencies. In this case, the model must break down as RMF penetration is imperfect at high densities and high collisionality. This must be an area of future exploration.

In the case of incompletely penetrated plasma, rigid rotation may only hold up to a certain radius, or equivalently up to a certain flux contour. Alternatively, the entire FRC may spin at the same rate regardless of RMF penetration.\cite{rostoker_equilibrium_2002} This case is not explored in this analysis, but one might expect a piecewise flux function in this case, where fluxes smaller than the penetration flux have the dependence given in Equation \ref{eq:pfrompsi} or \ref{eq:pexp}, and fluxes larger than the penetration flux have some shallower dependence.

One could decouple the synchronous RMF frequency and the RF frequency by using a a high-azimuthal-mode-number RMF antenna. Azimuthal mode number refers to $\vec{B}_{\hat{r},\hat{\phi},RMF}\propto e^{im\phi}$, where $m$ is the azimuthal mode number. $m=\pm1$ is a straight field, which existing RMF antennae produce. $m=\pm2$ would be a quadrupole field. $|m|>1$ fields vanish at $r=0$. It takes a point of constant RMF phase $m$ RF periods to make one revolution of $\phi$, so the RMF-synchronous frequency is a factor of $m$ smaller than the RF frequency.

\begin{figure}
\includegraphics[width=0.45\textwidth]{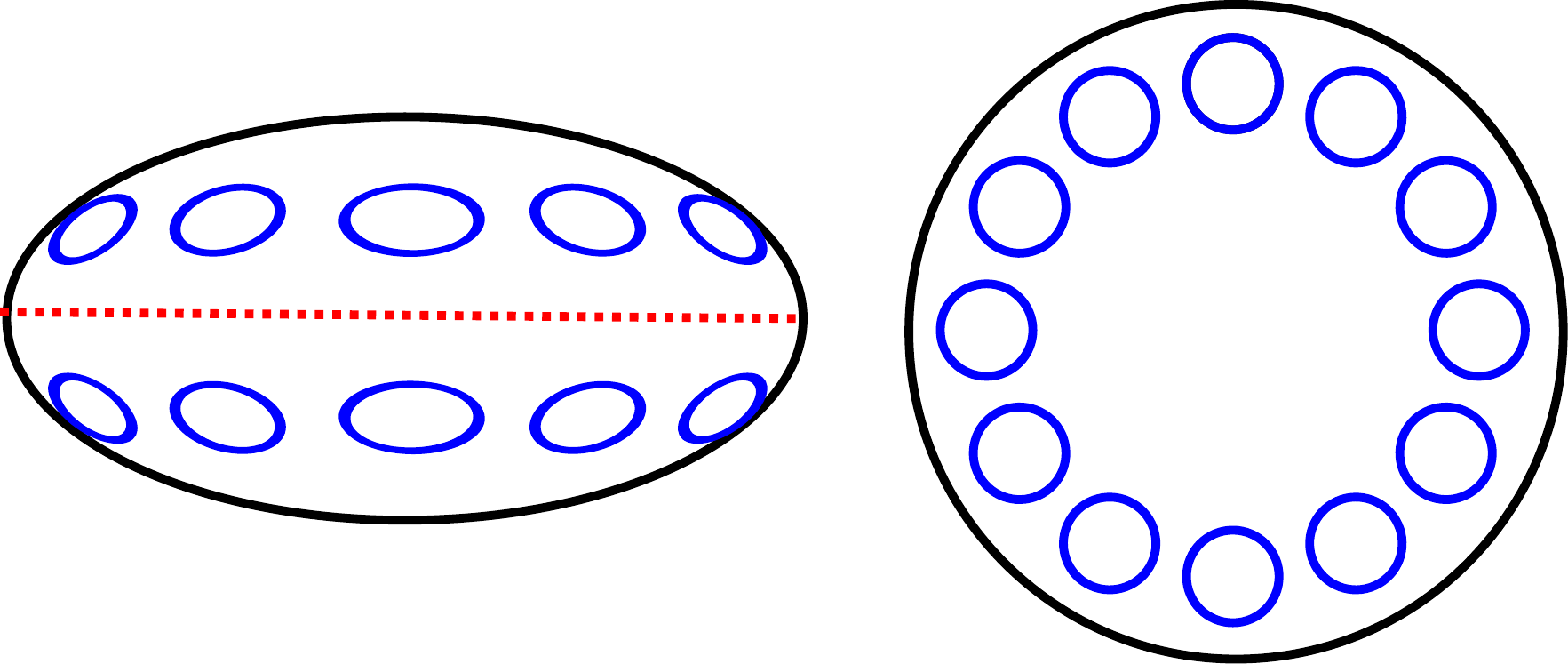}
\caption{\label{fig:oblatevessel} Location of odd-parity RMF antennae leaning into the oblateness of the FRC. Black: Vacuum vessel. Red dashed line: Vessel equator. Blue circles: RMF antennae. Left: Side view, $\hat{z}$-axis points upward. Right: Top view, $\hat{z}$-axis points out of the page. }
\end{figure}

\begin{figure}
\includegraphics[width=0.5\textwidth]{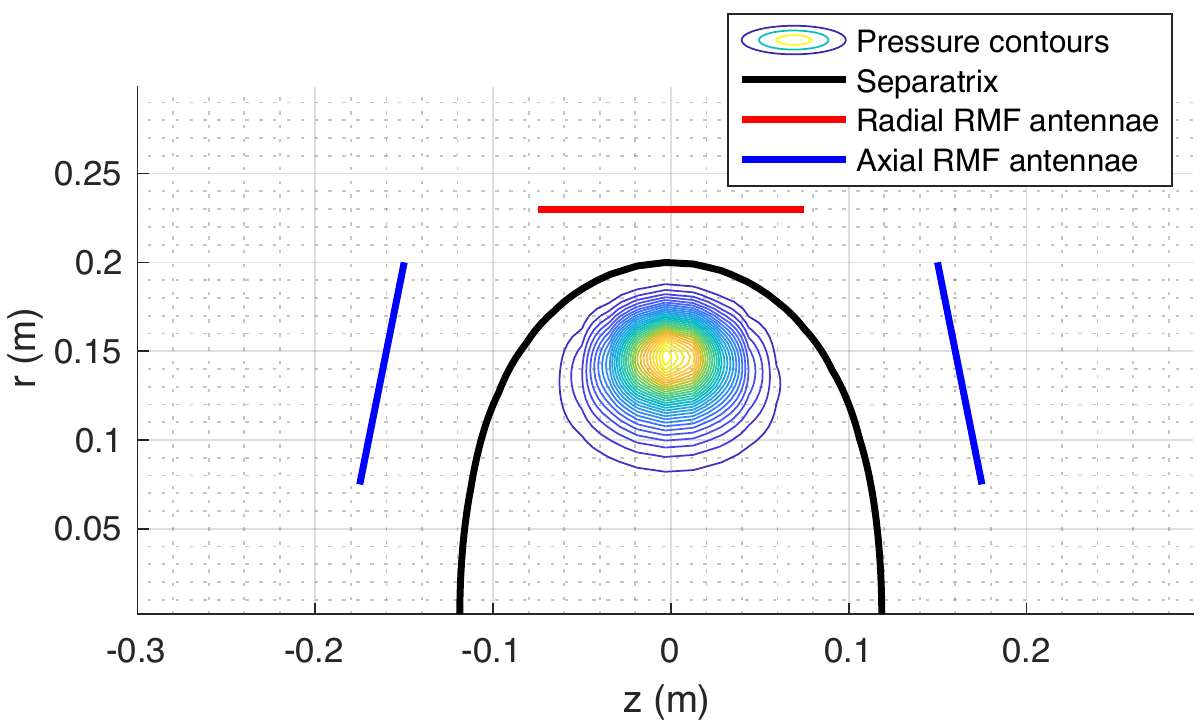}
\caption{\label{fig:oblateantennae} An equilibrium with $N=3$ showing an oblate separatrix and peaked pressure profile, revealed by localized pressure contours. Red: The location of a traditional RMF antenna, even- or odd-parity, producing a radial magnetic field. Blue: The location of RMF antennae leaning into the natural oblateness of the FRC, producing axial magnetic fields. These are naturally odd-parity if the antennae have the same axial polarity. The axial RMF antennae are shown tilted to show that they can have some radial component and still produce useful and odd-parity fields. }
\end{figure}

An intriguing possibility arises from leaning into the natural oblateness of FRCs with peaked spatial density profiles. This may be a novel and interesting parameter regime for later study. Rotating magnetic field coils facing axially, or a combination of axially and radially, rather than radially are naturally odd-parity, see Figures \ref{fig:oblatevessel} and \ref{fig:oblateantennae}, and the lower aspect ratio of this configuration gives plenty of space to include antennae at multiple azimuthal locations for high-mode-number RMF antennae. The coils would be roughly circular, flush with the oblate vacuum vessel, and spaced azimuthally. This would allow the radial profile of the RMF to be tailored. It would also allow a decoupling of the applied RF and synchronous rigid rotation frequencies. 

\begin{acknowledgments}
We wish to acknowledge Bruce Berlinger for his work on the hardware of the PFRC-2 experiment. We wish to acknowledge graduate students Eugene Evans and George Constantinos for helpful discussions. We acknowledge Stephanie Thomas and Michael Paluszek for contributing to the PFS Fusion Energy Toolbox.

This work was supported by DOE contract DE-AR0001099, and in part by the Program in Plasma Science and Technology, DOE contract DE-AC02-09CH11466, NASA contract 80NSSC18C0218, and NASA contract 80NSSC18K0040.
\end{acknowledgments}

%\nocite{*}
\bibliography{RMFGradShafranov}% Produces the bibliography via BibTeX.

\end{document}